\documentclass[runningheads]{llncs}
\usepackage[utf8]{inputenc}
\usepackage[hidelinks,bookmarks=true,bookmarksopen=true]{hyperref}
\usepackage{graphicx}
\usepackage{eurosym}
\usepackage{color}

\begin{document}

\title{DXP: Billing Data Preparation\\for Big Data Analytics}
%
%

\author{Luca Gagliardelli \and Domenico Beneventano \and Marco Esposito$^\dagger$ \and Luca Zecchini \and Giovanni Simonini \and Sonia Bergamaschi \and Fabio Miselli$^\diamond$ \and Giuseppe Miano$^\diamond$}
\authorrunning{L. Gagliardelli et al.}
%
\institute{University of Modena and Reggio Emilia, $^\diamond$Doxee S.p.A., Modena, Italy\\
\email{\texttt{\{name.surname\}@unimore.it} $^\dagger$\texttt{279958@studenti.unimore.it}\\
$^\diamond$\texttt{\{fmiselli,gmiano\}@doxee.com}}
}

\maketitle              

\begin{abstract}
In this paper, we present the data preparation activities that we performed for the Digital Experience Platform (\texttt{DXP}) project, commissioned and supervised by Doxee S.p.A..
\texttt{DXP} manages the billing data of the \emph{users} of different \emph{companies} 
operating in various sectors (electricity and gas, telephony, pay TV, etc.).
This data has to be processed to provide services to the users (e.g., interactive billing), but mainly to provide analytics to the companies (e.g., churn prediction or user segmentation).
We focus on the design of the data preparation pipeline, describing the challenges that we had to overcome in order to get the billing data ready to perform analysis on it.
We illustrate the lessons learned by highlighting the key points that could be transferred to similar projects.
Moreover, we report some interesting results and considerations derived from the preliminary analysis of the prepared data, also pointing out some possible future directions for the ongoing project, spacing from big data integration to privacy-preserving temporal record linkage.

\end{abstract}

\section{Introduction}
Data quality represents one of the main goals of every (big) data integration process and constitutes an essential requirement to get consistent and useful results when operating on the obtained integrated data.
This is particularly evident for data analysis, which considers the integrated data as a starting point.
Of course, performing analysis on dirty data can only lead to poor-quality and potentially wrong results (\emph{garbage in, garbage out}), impacting on the goodness of the choices deriving from data-driven decision-making processes.
Furthermore, with the increasing adoption of machine learning and deep learning models for big data analysis, it is of primary importance to ensure the quality of the data these models are trained on.
Recent developments in research community aim at moving from a model-centric to a \emph{data-centric} approach to artificial intelligence\footnote{\url{https://datacentricai.org}}, putting the emphasis on data quality rather than on model refinement.
In fact, even state-of-the-art models fail to provide good results if the data is not properly prepared \cite{DBLP:conf/vldb/ZecchiniSB20}. 
Data preparation and cleaning often represent the heaviest burden for the data scientist.
This phases include a wide range of operations, generally combined into pipelines, which can only be partially automated \cite{DBLP:journals/sigmod/HameedN20}.

The ongoing Digital Experience Platform (\texttt{DXP}) project
was commissioned by Doxee S.p.A.\footnote{\url{https://www.doxee.com}} to the Database Research Group (DBGroup\footnote{\url{https://dbgroup.unimore.it}}) of the University of Modena and Reggio Emilia in 2020, to leverage its experience in big data integration and analysis \cite{DBLP:books/sp/18/BergamaschiBMM0OPVSZGM18}, even in industrial scenarios \cite{gagliardelli2022ecdp,gagliardelli2018bigdedup}.
The goal of \texttt{DXP} is to acquire the billing data provided by the suppliers (\emph{companies}), then to process it to produce analytics for these companies (e.g., churn prediction or customer segmentation) and reports for their \emph{users} (e.g., interactive billing).
As highlighted in Section~\ref{sec:data_preparation}, the acquired data needs an intensive preparation to get ready to be used in data analysis.
In this paper, we illustrate the data preparation pipeline conceived for the project, together with an overview of the preliminary analysis of the prepared data (providing some considerations about the main points that we found out) and some future directions for the project, concerning big data integration and privacy-preserving temporal record linkage issues.
We focus on lessons learned and on reproducible aspects of the project, aiming at transmitting to the reader a method that can be adopted to make these time-consuming steps easier while dealing with data analysis tasks.

\section{Data Preparation Pipeline}
\label{sec:data_preparation}
The first step needed to obtain significant analytics from the available billing data was the design of a proper data preparation and cleaning pipeline.
This pipeline was required to be general purpose, to be employed for the billing data of the different companies managed by Doxee.

During the starting phase of the project, we worked on a subset of about ten million bills of a single company, covering a period of six months (from January to June 2021).
Each bill is represented by a JSON file that is used by Doxee to provide the users with an electronic bill that can be accessed online.
The JSON files (one for each bill) are organized in folders (one for each month).
Each JSON file contains three types of information: \emph{(i)} user data; \emph{(ii)} Point Of Delivery (POD) data; \emph{(iii)} specific bill information.
In the considered JSON files provided for the project, the personal data about the user are anonymized through a hashing function to be GDPR compliant. Since each value is represented by the same hash value, it is therefore possible to design the data preparation pipeline without accessing the personal information.

\begin{figure}[hb]
    \centering
    \includegraphics[width=\linewidth]{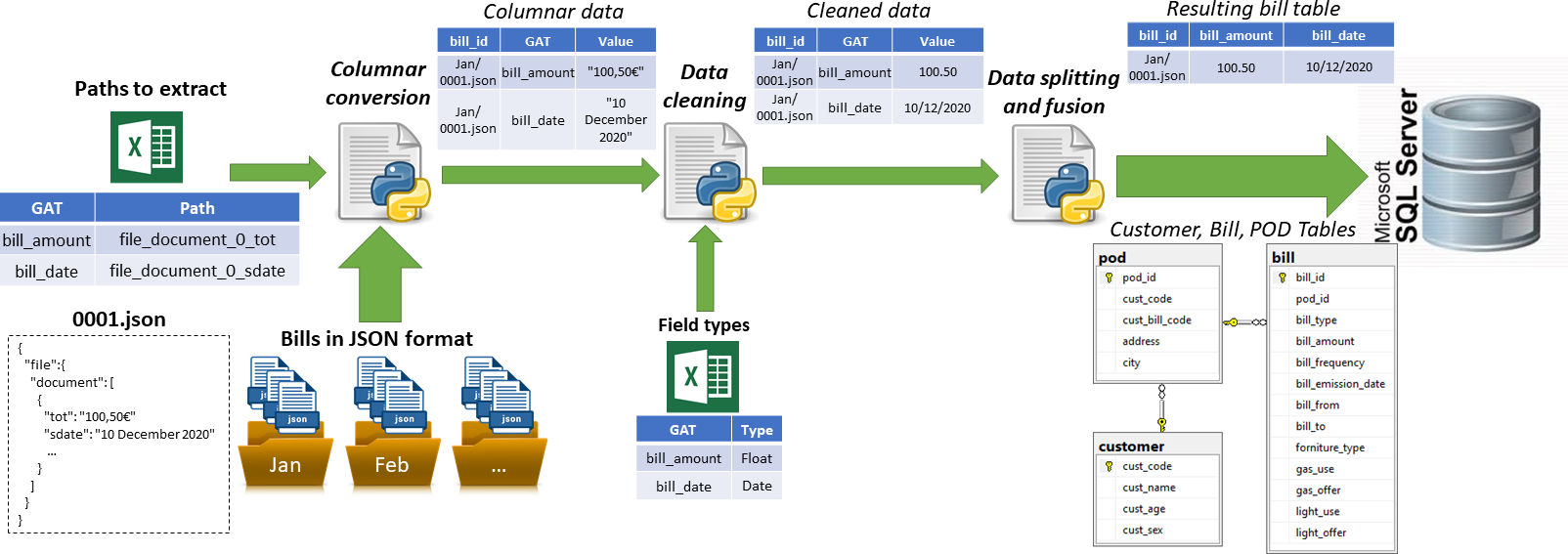}
    \caption{Data preparation pipeline. The table ``Bill" reports the most relevant fields.}
    \label{fig:pipeline}
\end{figure}

The pipeline accepts as input the JSON files, producing as output three relational tables that contain the cleaned data about the bills, the PODs, and the users.
The relational tables are stored into an instance of Microsoft SQL Server, since Microsoft Power BI, chosen to perform the data analysis, can query it seamlessly. 
The pipeline, composed of three phases, is depicted in Figure~\ref{fig:pipeline} and was developed with Python by using Apache Spark to be deployed on cloud services.

The most challenging step was to identify the attributes to be extracted from the JSON files.
In fact, these files have high complexity and contain a lot of information which is used to generate the interactive electronic bills.
To overcome this issue, we used a sample bill with the related JSON file to identify the path (or the paths) corresponding to each attribute.
Moreover, a domain expert helped us to identify the fields of interest and the ones used as unique identifiers for each POD.
To make the process more flexible, we defined a spreadsheet that for each attribute of interest (Global Attribute, GAT) to extract defines the path where the attribute is located in the JSON file.
Thus, if a new attribute needs to be extracted, it just has to be added to this file.

The first step of the pipeline takes as input the spreadsheet with the paths and the folders that contain the JSON files, producing as output a columnar file that for each bill and each GAT reports the corresponding value contained in the bill.
To identify each bill, we chose to use its full path.
This makes it easier to debug the process in case of extraction problems, since we can immediately find the JSON file from which the values were extracted.

The second step of the pipeline is devised to clean the data.
In fact, the data extracted from the JSON files presents many issues, being formatted for viewing purposes.
For example, numerical data contains measure units to be removed and dots/commas to be normalized according to a common standard, in order to store it in a numerical format (e.g., the total amount of a bill is stored as ``1.000,00 \euro"; thus, it has to be transformed into ``1000.00" to be converted into a floating-point format).
Furthermore, the dates are written as ``\textit{day month year}" (e.g., ``10 January 2021") and need to be converted into date format.
To face these issues, we prepared a set of custom functions to be applied to the value of each field according to the expected output type defined in a spreadsheet.
This makes it easier to perform future changes, such as adding new attributes to the extraction pipeline.
At the end of this step, we obtain a clean version of the data.

\begin{figure}[t]
    \centering
    \includegraphics[width=0.45\linewidth]{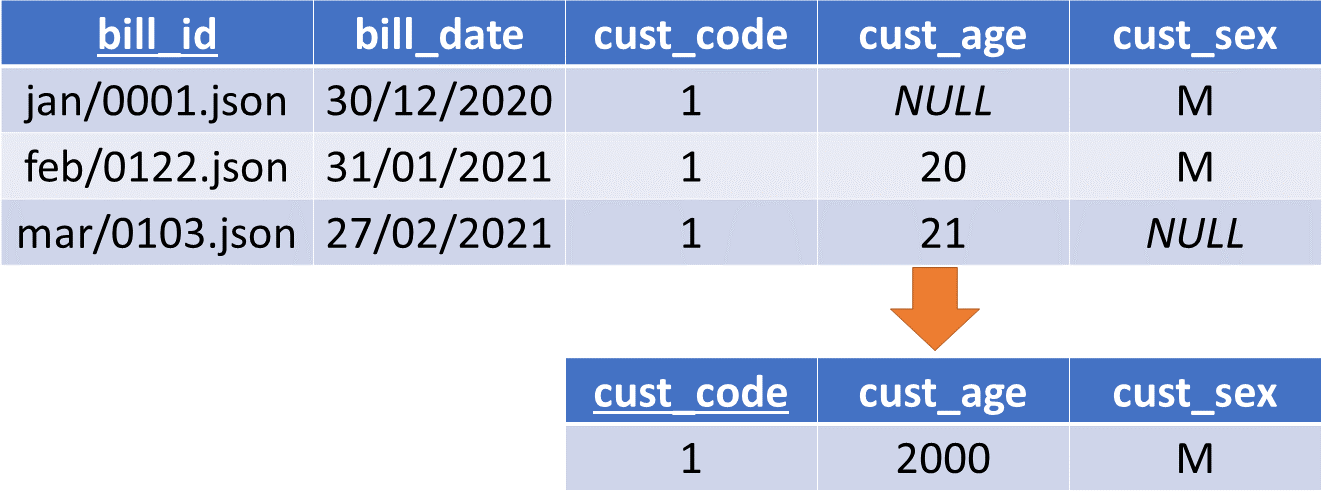}
    \caption{Data fusion.}
    \label{fig:row_err}
\end{figure}

The third and final step of the pipeline aims at generating the relational tables.
First, a pivot operation is performed on the data to obtain a table which contains a record for each bill.
The resulting table is denormalized: the data about a POD is repeated as many times as there are bills associated with it, while the data about a user is repeated as many times as there are PODs associated with them.
For example, if a user owns two PODs and for each POD three bills were emitted, the data about that user is repeated six times.
This representation presents several problems: \emph{(i)} it is inefficient in terms of memory occupation, since the same values are repeated multiple times; \emph{(ii)} it is inefficient to be queried (e.g., to find all the bills concerning the same POD it is necessary to scan the whole table, due to the lack of a foreign key referring to the POD); \emph{(iii)} finally, there can be inconsistencies among the different records that refer to the same user/POD.
An example is shown in Figure~\ref{fig:row_err}, where for the same user the sex is missing in a bill and the age changes from one bill to another.
Then, after the pivoting, the fields are divided into three different relational tables according to what they refer to (user, POD, or bill), as shown in Figure~\ref{fig:pipeline}.
Finally, to overcome the inconsistencies, we adopted a resolution function that takes for each POD/user the most recent not null value for every attribute, computing the year of birth and considering it instead of the age, as shown in Figure~\ref{fig:row_err}: the function takes ``2000" for the age and `M' for the sex because they are the most recent not null values according to the date of the bill.

\section{Data Analysis}
\label{sec:data_analysis}
In this preliminary data analysis phase, we focused only on the electricity bills.
For these bills, each POD is associated to an offer that determines the cost of the electricity.
The offer is reported in each bill, and a user can decide to leave an offer for another one (e.g., to save money).
In some cases, the users may even decide to leave the current electricity supplier, choosing an offer by a different company.
The goal of our analysis was to perform a churn prediction, in order to forecast when a user may be interested in leaving the current offer, allowing the company to suggest a different one providing economic benefits, hopefully preventing the switching to a different supplier.
We decided to do this by using a classification model.

The first challenge was to construct a feature vector to feed the model.
The prediction task is related to each POD and to each offer present in the bills for that POD, so we had to create a feature vector depending on both elements, starting from the data contained in each bill.
For each POD-offer pair, we chose the following features: data of the user who owns the POD (i.e., \emph{sex} and \emph{age}), \emph{municipality} where the POD is located, data of the bills related to that POD and that offer (\emph{total consumption}, \emph{total amount}, \emph{total light amount}, \emph{billed days}), and \emph{churn}.
The \emph{total consumption} is expressed in kilowatts, the \emph{total light amount} denotes the cost of the consumed electricity, while the \emph{total amount} represents the full cost, including extra fees.
The \textit{churn} is the target label (i.e., the variable to predict): it is set to 1 (i.e., the POD left the offer) if the last offer associated with that POD (i.e., the offer in the last available bill) is different from the offer that appears in the feature vector; otherwise, it is set to 0 (i.e., the POD maintained that offer).
All textual attributes (\emph{offer}, \emph{sex}, \emph{municipality}) were transformed into numerical by using a categorical encoding, in order to be processed by the classification model.
We performed a correlation analysis on the features through Pearson correlation coefficient, finding a scarce correlation between them and the target variable.
However, we found a weak inverse correlation between the number of billed days and the churn, i.e., some users frequently change the offer to obtain better economic conditions.

Finally, we chose \emph{random forest} as a classification model to enhance the accuracy \cite{DBLP:journals/jcst/LiWLKLZL15} (we also tried \emph{logistic regression} and \emph{SVM}), then we performed a cross-validation test to evaluate the performance of the model, which gained an accuracy of $98.7\%$, but presenting an unsatisfactory recall of $51\%$ on the users who changed the offer (i.e., \textit{churn} = 1).
This is mainly due to the high imbalance between the two classes present in the available data, since only $1.8\%$ of the records is related to users who changed their offer.
We also performed stratification to balance the training data, obtaining similar results.

The conclusion of this preliminary analysis was that more data is needed for this step.
In particular, the available data presents a lack of temporal depth: since for each POD a bill is usually emitted every two months, six months are not enough to obtain satisfactory results.
As a general principle, it is preferable to focus on the complete history of a significant and representative subset of users, since the study of these time series allows to understand the behavior of these users (or types of users), and in particular how it changes over time, detecting relevant patterns to train a significant model.

\section{Conclusion and Future Work}
In this paper, we described our experience in tackling a real-world project aimed at big data analysis, highlighting the importance of data preparation in order to obtain correct and significant results.
In particular, our paper is focused on the design of a data preparation pipeline, conceived dealing with the billing data of a specific company referring to a defined time interval, but with the purpose to be reusable for future billing data and generalizable to the billing data of any other company assisted by Doxee.
We also report some considerations derived from our preliminary analysis about the characteristics that input billing data needs to satisfy to allow the extraction of significant information from it.

We described the data preparation pipeline and the preliminary data analysis referring to the billing data of a single company. On the other hand, a key aspect of \texttt{DXP} that will play a central role in its future developments is the analysis of the integrated data about the users, provided by the different companies. As user data contains private and confidential information, 
the impact of privacy requirements and GDPR compliance on big data integration must be taken into account. The data integration techniques developed by the DBGroup will be extended in this direction; in particular, our entity resolution and blocking solutions will be enhanced to consider the \emph{privacy-preserving temporal record linkage} scenario \cite{vatsalan2017privacy}.

\section*{Acknowledgements}
We would like to thank Marcello Generali and Monia Gazzano from Doxee S.p.A. for their supervision and contributions.
The project Digital Experience Platform (\texttt{DXP}) was funded by Regione Emilia-Romagna as part of the program for the promotion of companies' development\footnote{\url{https://bur.regione.emilia-romagna.it/dettaglio-inserzione?i=280032d4b5e7426b98eb92d955149e72}}.

\bibliographystyle{splncs04}
\bibliography{main}

\begin{thebibliography}{1}
\providecommand{\url}[1]{\texttt{#1}}
\providecommand{\urlprefix}{URL }
\providecommand{\doi}[1]{https://doi.org/#1}

\bibitem{DBLP:books/sp/18/BergamaschiBMM0OPVSZGM18}
Bergamaschi, S., Beneventano, D., Mandreoli, F., et~al.: {From Data Integration
  to Big Data Integration}. In: A Comprehensive Guide Through the Italian
  Database Research Over the Last 25 Years, Studies in Big Data, vol.~31, pp.
  43--59. Springer (2018)

\bibitem{gagliardelli2022ecdp}
Gagliardelli, L., Zecchini, L., Beneventano, D., et~al.: {ECDP: A Big Data
  Platform for the Smart Monitoring of Local Energy Communities}. In: DataPlat.
  {CEUR} Workshop Proceedings, vol.~3135. CEUR-WS.org (2022)

\bibitem{gagliardelli2018bigdedup}
Gagliardelli, L., Zhu, S., Simonini, G., Bergamaschi, S.: {BigDedup: A Big Data
  Integration Toolkit for Duplicate Detection in Industrial Scenarios}. In: TE.
  Advances in Transdisciplinary Engineering, vol.~7, pp. 1015--1023. IOS Press
  (2018)

\bibitem{DBLP:journals/sigmod/HameedN20}
Hameed, M., Naumann, F.: {Data Preparation: {A} Survey of Commercial Tools}.
  {SIGMOD} Rec.  \textbf{49}(3),  18--29 (2020)

\bibitem{DBLP:journals/jcst/LiWLKLZL15}
Li, H., Wu, D., Li, G., Ke, Y., Liu, W., Zheng, Y., Lin, X.: {Enhancing Telco
  Service Quality with Big Data Enabled Churn Analysis: Infrastructure, Model,
  and Deployment}. J. Comput. Sci. Technol.  \textbf{30}(6),  1201--1214 (2015)

\bibitem{vatsalan2017privacy}
Vatsalan, D., Sehili, Z., Christen, P., Rahm, E.: {Privacy-Preserving Record
  Linkage for Big Data: Current Approaches and Research Challenges}. In:
  Handbook of Big Data Technologies, pp. 851--895. Springer (2017)

\bibitem{DBLP:conf/vldb/ZecchiniSB20}
Zecchini, L., Simonini, G., Bergamaschi, S.: {Entity Resolution on Camera
  Records without Machine Learning}. In: DI2KG. {CEUR} Workshop Proceedings,
  vol.~2726. CEUR-WS.org (2020)

\end{thebibliography}

\end{document}